\documentclass[aps,twocolumn,secnumarabic,amsmath,amssymb,nobalancelastpage,nofootinbib,natbib,preprintnumbers]{revtex4-1}
\pdfoutput=1
\usepackage[
      colorlinks=true,
      linkcolor=blue,
      urlcolor=blue,
      filecolor=black,
      citecolor=red,
      pdfstartview=FitV,
      pdftitle={},
        pdfauthor={Paulo Bedaque, Evan Berkowitz, Geoffrey Ji, Nathan Ng},
        pdfsubject={},
        pdfkeywords={},
        pdfpagemode=None,
        bookmarksopen=true
      ]{hyperref}
      
\usepackage{amsmath, bm, graphicx}
\usepackage{epstopdf}
\usepackage{amscd}
\usepackage{slashed}

\newcommand{\figref}[1]{Fig.~\ref{#1}}

\newcommand{\CFLKz}{CFL-\Kzero}
\newcommand{\Kplus}{{\ensuremath{K^+}}}
\newcommand{\Kminus}{{\ensuremath{K^-}}}
\newcommand{\Kzero}{{\ensuremath{K^0}}}

\newcommand{\fpi}{\ensuremath{f }}
\newcommand{\f}{\fpi}
\newcommand{\pF}{\ensuremath{p_F}}

\newcommand{\One}{\ensuremath{\openone}}
\newcommand{\M}{\ensuremath{\mathbb{M}}}
\newcommand{\Q}{\ensuremath{\mathbb{Q}}}

\newcommand{\half}{\ensuremath{\frac{1}{2}}}

\newcommand{\Lge}{\ensuremath{\mathcal{L}_{\text{eff}}}}

\newcommand{\tension}{\ensuremath{\mathcal{F}_\perp}}

\newcommand{\adjoint}{\ensuremath{{}^\dagger}}
\newcommand{\conjugate}{\ensuremath{{}^*}}
\newcommand{\magnitude}[1]{\ensuremath{\left|#1\right|}}
\newcommand{\del}{\ensuremath{\nabla}}
\newcommand{\grad}{\del}
\newcommand{\orderof}[1]{\ensuremath{\mathcal{O}\left(#1\right)}}

\newcommand{\Tr}[1]{\ensuremath{\text{Tr}\left( #1 \right)}}
\newcommand{\Det}[1]{\ensuremath{\text{Det}\left(#1 \right)}}
\newcommand{\diag}[1]{\ensuremath{\text{diag}\left(#1 \right)}}
\newcommand{\inverse}{\ensuremath{^{-1}}}
\newcommand{\oneover}[1]{\ensuremath{\frac{1}{#1}}}
\newcommand{\hc}{\text{h.c.}}
\newcommand{\commutator}[2]{\ensuremath{\left[#1,#2\right]}}
\newcommand{\expect}[1]{\ensuremath{\left\langle#1\right\rangle}}

\newcommand{\arccoth}{\ensuremath{\text{arccoth}}}
\newcommand{\csch}{\ensuremath{\text{csch}}}

\begin{document}

\title{Electron shielding of vortons in high-density quark matter}
\author{Paulo F. Bedaque$^{1}$}
\email{bedaque@umd.edu}
\author{Evan Berkowitz$^{1}$}
\email{evanb@umd.edu}
\author{Geoffrey Ji$^{2}$}
\email{gji@umd.edu}
\author{Nathan Ng$^{3}$}
\email{nang@mbhs.edu}
\affiliation{$^{1}$ Maryland Center for Fundamental Physics, Department of Physics,
University of Maryland, College Park, MD 20742-4111 \\
$^{2}$University of Maryland, College Park, MD 20742\\
$^{3}$Montgomery Blair High School, 51 University Boulevard East, Silver Spring, MD 20901
}
\preprint{UMD-DOE/ER/40762-511}

\begin{abstract}
We consider the the effect of the electron cloud about a vorton in the \CFLKz\ high-density phase by numerically solving the ultrarelativistic Thomas-Fermi equation about a toroidal charge. Including electrons removes the electric monopole contribution to the energy, and noticeably decreases the equilibrium radius of these stable vortex loops.
\end{abstract}

\maketitle
\section{Introduction} 
\label{sec:introduction}

The behavior of matter at very high densities, like those found at the core of neutron stars, remains  largely an open problem. This is due to a lack of reliable methods to extract information directly from QCD. But since the temperatures relevant for the physics of neutron stars is very low (as compared to the Fermi energy) a number of observables of  interest depend only on properties of the low lying excitations of the ground state. In the case of dense matter it is believed that those excitations are pseudo-Nambu-Goldstone (NG) bosons arising from the spontaneous breaking of certain approximate symmetries. The study of these quasiparticles using the methods of effective field theory opens up the possibility of a more reliable understanding of dense matter.

The breaking of symmetries in high-density QCD is related to quark pairing. At high density, evidence now points towards color superconductivity where quarks pair in different combinations depending on the temperature and chemical potential\cite{Barrois:1977xd,Bailin:1983bm,Alford:1997zt,Rapp:1997zu}.  This pairing uses the BCS mechanism, just like electron pairing in conventional electrical superconductors.  However, color superconductivity is a much richer phenomenon than electrical superconductivity because quarks carry flavor and color while electrons do not---different color superconducting phases arise depending on which quarks pair. The nature of the NG bosons depends crucially on the pattern of symmetry breaking and, consequently, on the kinds of quark pairing found in the ground state.

At asymptotically high density, three-flavor QCD is believed to pair in a pattern known as color-flavor-locking (CFL), as it pairs all quarks\cite{Alford:1998mk}.  At lower densities, where the QCD coupling is not small, the problem is more complicated and a plethora of  phases have been suggested (for a review see \cite{Alford:2001dt,Rajagopal:2000wf}).  Perhaps the most likely phase at very high density is the \CFLKz\ phase, which seems guaranteed to be the energetically favored phase for some range of large densities\cite{Bedaque:2001p107,Kaplan:2001qk}.  This range may include densities relevant for compact astrophysical objects and thus may be phenomenologically relevant for extracting predictions about extreme environments, like the inner cores of neutron stars.  

The NG modes of the CFL phase form an SU(3) octet, and are named in direct analogy with the \emph{in vacuo} SU(3) flavor octet, but at high density the ``kaons'' are the lightest particles \cite{Son:1999cm,Beane:2000ms,Son:2000tu,Casalbuoni:1999wu,Casalbuoni:1996pg}.  As they live in an SU(3) octet, their behavior is described by a theory resembling familiar chiral perturbation theory.  As the chemical potential $\mu$ is lowered and the quark masses become relevant some of the NG modes become lighter.  When $\mu$ is low enough, the neutral ``kaons'' would have a negative mass-squared, and instead condense with some vacuum expectation value, turning the CFL phase into \CFLKz \cite{Bedaque:2001p107,Kaplan:2001qk}.  This neutral kaon condensate provides a superfluid background that is the stage for interesting topological defects. The reason the neutral, as opposed to the charged, kaons condense is the small differences in their mass induced by isospin violating quark masses and electromagnetic effects. These differences are small so a condensate of charged kaons is almost degenerate with  the favored neutral kaon condensate.  This peculiar situation provides a physical realization of the so-called ``superconducting string" solutions, suggested in more abstract contexts\cite{Witten:1984eb} and also appearing in some models describing beyond the Standard Model physics\cite{Hindmarsh:1994re,vilenkin2000cosmic,Davis1989209,Davis:1988ij} . Regular global vortices appear due to the \Kminus condensation. At their core, the value of the \Kzero\ condensate vanishes, as it is the case in all vortices. Since there is a close competition between neutral and charged kaon condensates, the charged kaons will then condense in the core of the \Kzero\ vortex. The condensation of a charged particle leads to \emph{electrical} superconductivity (not color superconductivity) along the vortex. Loops of superconducting vortices (``vortons") can be stabilized by the supercurrent running around them. The mechanism, roughly speaking, is that the energy of a vorton has two main contributions. One is the  tension along their length, which creates an energy that scales with its radius $R$ and drives a vorton to be small.  However, the energy of a vorton includes terms that scale like  $J^{2}/Q^{2}R$ where  $J$ is its angular momentum and $Q$ its electric charge.  These terms prevent a vorton from collapsing. This mechanism was invoked in \cite{Kaplan:2001hh} in the context of quark dense matter. It was subsequently pointed out \cite{Buckley:2002ur,Buckley:2002p109}  that, just like in the case of cosmological vortons, the currents necessary to stabilize a vorton could be high enough that they would quench the superconductivity of the vortex core, destroying its supercurrent and rendering the stability mechanism mute. The quenching effect can, however, be counteracted by electric charge. A vorton with an overall electric charge, in addition, would be more stable on account of Coulomb repulsion between opposong sides of the loop.

A more quantitative theory of quark matter vortons appeared only recently. In Ref.~\cite{Bedaque:2011fg}, the equilibrium size of vortons containing charge as well current was estimated by making a variety of approximations and assumptions.  It was found that stabilizing a vorton so that its equilibrium radius was several times the thickness of its core $\delta$ and thus could really be identified as a torus, required large electric charges ($Z\sim10^{4}$).  With such a large charge, it is possible that ambient electrons which are in the bulk to ensure charge neutrality could be pulled close to the vorton, shielding its electric fields, and significantly reducing the effectiveness with which the electric charge aided in stabilizing the vorton in the first place.  We call these vortons with neutralizing orbiting electrons vortonium to easily distinguish them from unshielded or bare vortons.

It is the purpose of this paper to account for these electrons and to find the equilibrium radius of a vortonium ``nucleus''.  We use the Thomas-Fermi approximation to find the electrons' semiclassical charge distribution, allowing them to shield the electric field.  Second, having found the electron distribution, we can compute the energy in the electric field numerically, without making geometrical or multipole approximations.  Intuitively, when including electrons we should find that vortons have a smaller equilibrium radius, because there are exactly enough electrons to kill the monopole field.  We find that making these  improvements decreases the equilibrium radius of a vorton of given angular momentum and electric charge significantly.
 
\section{CFL+\Kzero} 
\label{sec:cfl_kzero}

To understand vortons, we first must understand the phase in which they appear.  When the quark masses are negligible compared to the chemical potential $\mu$, the CFL phase is described by the quarks pairing according to\cite{Schafer2000269}
\begin{equation}
	\expect{q^{a}_{L,i}C q^{b}_{L,j}}= - \expect{q^{a}_{R,i}C q^{b}_{R,j}} = \oneover{g_{s}}\mu^{2}\Delta\	\epsilon^{abZ}\epsilon_{ijZ}
\end{equation}
where $q$ are the quarks which carry color $a,b\in\{1,2,3\}$, flavor $i,j\in\{u,d,s\}$, and helicity $R,L$ indices.  The dimensionful scale $\Delta$ is known as the gap, and $g_{s}$ is the strong coupling constant.  This condensate breaks the full $SU(3)_{C}\times SU(3)_{L}\times SU(3)_{R}\times U(1)_{B} \times U(1)_{A}$ symmetry down to $SU(3)_{C+L+R}\times \mathbb{Z}_{2} \times \mathbb{Z}_{2}$, where the first remaining $\mathbb{Z}_{2}$ comes from breaking $U(1)_{B}$ and the second from breaking the approximate $U(1)_{A}$ symmetry that Dirac fermions enjoy at asymptotically high density.

Simple counting suggests that there should be 18 Nambu-Goldstone mesons, but eight of these are eaten by the gluons, which gain masses $\sim g_{s}\mu$, indicating that the phase is color superconductive\cite{Son:1999cm}.  The condensate breaks the electromagnetic $U(1)_{EM}\subset SU(3)_{L}\times SU(3)_{R}$, but there remains an unbroken gauge symmetry that arises from a linear combination of the eighth gluon and the photon.  This unbroken gauge symmetry, which we call $U(1)_{Q}$ is ``mostly'' electromagnetism: when $\alpha_{EM}$ is small the contribution from the eighth gluon is small, the differences between $\alpha_{Q}$ and $\alpha_{EM}$ are small, and the dielectric constant $\epsilon_{CFL}$ differs from $\epsilon_{vacuum}$ by only a small amount.  All of these corrections are at least \orderof{\alpha_{EM}} suppressed, and we henceforth refer to the remaining $U(1)_{Q}$ symmetry as electromagnetism for convenience.

After the gluons eat eight of them, ten NG modes remain.  Two describe bulk superfluidity that results from breaking the baryon and axial symmetries.  The other eight are an octet of the $SU(3)_{C+L+R}$.  Because these eight ``mesons'' share group structure with the \emph{in vacuo} mesons that are described by flavor SU(3), these modes are named in an analogous manner.  The theory contains a chiral field $\Sigma$ which can take on a vacuum expectation value $\Sigma_{0}$.  The eight mesons $\pi^{a}$ are small fluctuations about $\Sigma_{0}$, so that $\Sigma=\exp\left(	i\pi^{a}\lambda^{a}/f	\right)\Sigma_{0}$, where $\lambda^{a}$ are the usual 8 Gell-Mann matrices with $\Tr{\lambda^{i}\lambda^{j}}=2 \delta^{ij}$ and $f$ plays a role analogous to the pion decay constant.

To lowest-order, the effective theory describing the CFL phase is described by the Lagrangian~\cite{Son:1999cm,Beane:2000ms,Son:2000tu,Casalbuoni:1999wu,Bedaque:2001p107}
\begin{align}
	\Lge 	&=	\frac{\f^{2}}{4}\Tr{ \grad_{0}\Sigma\adjoint\cdot\grad_{0}{}\Sigma - v^{2}D_{i}\Sigma\adjoint\cdot D_{i}\Sigma}\nonumber\\
			&\phantom{=\ }+2 A \Det{\M}\Tr{\M\inverse \Sigma + \hc} - \oneover{4}F^{2}	\label{eq:lagrangian} ,
\end{align}
where the derivatives are
\begin{align}
	D_{\mu}\Sigma 	&= 	\partial_{\mu}\Sigma - i A_{\mu}\commutator{\Q}{\Sigma},	\nonumber\\
	\grad_{0}\Sigma &=	D_{0}\Sigma + i\commutator{\frac{\M\M\adjoint}{2 \pF}}{\Sigma}, \label{eq:CovariantDerivatives}
\end{align}
$\M = \diag{m_{u},m_{d},m_{s}}$ is the quark mass matrix, $\Q=\frac{e}{3}\diag{2,-1,-1}$ is the matrix of the quark charges under $U(1)_Q$, $A_{\mu}$ is the $U(1)_Q$ gauge field, $F$ its corresponding field strength tensor, $A$, $f$, and $v$ low-energy constants, and $\pF$ the Fermi momentum, which we take to be equal to the chemical potential $\mu$.  

The leading-order mass term differs from the usual chiral Lagrangian, which has a term linear in \M.  A heuristic way to understand this difference is that the CFL condensate leaves $\mathbb{Z}_{2}$ remnants of the left- and right-$SU(3)$ symmetries, so that left- and right-handed quark number should be preserved modulo 2, while terms linear in \M\ mix left- and right-handedness and violate this conservation and are thus forbidden.  The mass term included in this Lagrangian is not the only \orderof{\M^2} term, but the other possible terms are higher-order in the power-counting scheme developed in Ref.~\cite{Bedaque:2001p107}.

Interestingly, the low-energy constants can be matched to perturbative QCD calculations at asymptotic $\mu$.  Such matching calculations give~\cite{Son:1999cm,Son:2000tu}
\begin{equation}
	\f^{2}=\frac{21-8\ln2}{18}\frac{\mu^{2}}{2\pi^{2}},\hspace{.75em}v^{2}=\oneover{3},\hspace{.75em}\text{and}\hspace{.5em}A=\frac{3}{4\pi^{2}}\Delta^{2}.
\end{equation}
Armed with this information, one can easily find the masses of the CFL mesons at asymptotically large $\mu$.  Expanding about $\Sigma_{0}=\One$, one finds that the \orderof{\M^2} term gives masses~\cite{Bedaque:2001p107}
\begin{align}\label{eq:meson-masses}
	m_{\pi^{\pm}} 	&= \mp \mu_{du}+	\sqrt{\frac{4A}{\f^{2}}(m_{d}+m_{u})m_{s}}			,\nonumber\\
	m_{K^{\pm}}		&=\mp \mu_{su}+	\sqrt{\frac{4A}{\f^{2}}m_{d}(m_{s}+m_{u})}			,\\
	m_{K^{0},\bar{K^{0}}}	&=\mp \mu_{sd}+	\sqrt{\frac{4A}{\f^{2}}m_{u}(m_{s}+m_{d})}	,\nonumber	
\end{align}
where $\mu_{ij} = (m_{i}^{2}-m_{j}^{2})/(2\mu)$ is the fictitious chemical potential which arises from the $\M$ commutator terms in the time-like derivatives.
As $\mu$ comes down from infinity, the kaons and charged pions get a fictitious strangeness chemical potential from the $\M\M\conjugate/2\pF$ commutator terms in the time-like derivatives, and are incentivized to condense.  The other mesons have even larger masses, but as they lie on the diagonal of $\Sigma$ when the other fields vanish, their masses do not get contributions from the commutator term and are not encouraged to condense.

Since the neutral kaons are the lightest, they condense first.  The charged kaons are have almost the same mass as the neutral kaons---the difference is dependent on the mass difference $m_{d}-m_{u}$ and also on electromagnetic effects.  However, the interactions between mesons in the effective theory strongly disfavors two meson species from condensing at once and, if \Kplus\ condensed, electrons would have to be sprinkled throughout the bulk with an additional cost in energy due to the electrons' Fermi energy.

With all of the other NG modes vanishing, the expectation value of the neutral kaons is~\cite{Bedaque:2001p107}
\begin{equation}
\label{eq:Kzero-soln}
	\cos\left(	\frac{\magnitude{\Kzero}\sqrt{2}}{\f}	\right) = \frac{m_{\Kzero}^{2}}{\mu_{sd}^{2}}
\end{equation}
where $m_{\Kzero}$ is given in \eqref{eq:meson-masses}.  Since cosine is restricted to be in the interval $[-1,1]$ we see that if $m_{\Kzero}^{2}/\mu_{sd}^{2}>1$, there is no neutral kaon condensate at all.  From the form of the chiral field $\Sigma = \exp\left(	i\pi^{a}\lambda^{a}/f	\right)\Sigma_{0}$, it is clear that the value of the kaon expectation value is a compact variable whose value only matters modulo $2\pi$, and thus it is not surprising that cosine shows up in the expression for the \Kzero\ condensate.

\section{Bare Vortons in \CFLKz} 
\label{sec:bare_vortons_in_cflkz}

The background condensate of neutral kaons provides a standard superfluid in which we can study global vortex strings.  As in any standard superfluid, the condensate must vanish at a vortex's center in order to remain single-valued.  One feature of the \CFLKz\ superfluid is that the other mesons could condense but interactions prevent more than one species from condensing.  Since the center of a superfluid vortex is evacuated, it is natural to expect that the cores of these vortices may contain a condensate of the next lightest species---the charged kaon.  Thus, the superfluid vortices have at their core a superconducting condensate which may carry electrical current or hold an electrical charge.  A vorton is a vortex loop with this superconducting core.

A detailed analysis of the stability of a vorton without electrons is given in Ref.~\cite{Bedaque:2011fg}.  There, an in-depth discussion about the stability of vortons with a fixed electrical charge and angular momentum is provided.  A vortex naturally has an associated tension, so a vortex loop would naturally shrink to nothing and destroy itself very quickly unless stabilized in some manner.  When the inner condensate does not break a gauge symmetry, the stability of a vorton is easy to understand.  Requiring that the core contain some fixed number of particles which have a preferred density can provide a pressure along the length of the vortex.  Alternatively, the condensate can have a phase which wraps around the length of the vortex.  Because the condensate must be single valued, this phase must be quantized, and is inherently topological.  The topological winding number $N$ enters quadratically into the tension \tension\ as $(N/R)^{2}$, while the usual terms in the tension are a constant $c$ independent of $R$ (the energy cost of suppressing the background condensate minus the savings provided by allowing the core to condense).  Then the energy $F\sim 2\pi R \tension \sim N^{2}/R + c R$ has a preferred equilibrium radius $R_{0}$.

However, the gauged case is a bit more tricky.  It was also noted in \cite{Bedaque:2011fg} that due to gauge invariance, the topological winding number $N$ of \Kplus\ that wraps around the length of the vorton and the magnetic fluxoid threaded through a vorton $\Phi_{B}$, while separate observables, always enter the Lagrangian and the energy functional $F$ in just the right combination so as to give the angular momentum $J$, so that we need not specify them separately but instead need only specify $J$.  In field theory, angular momentum depends on one space and one time derivative, while the charge depends on a time derivative.  From the form of the derivatives \eqref{eq:lagrangian} one concludes that the time and space derivatives are traded in the energy for $Q$ and $J/QR$.

In order to get an analytical feel for the energy functional $F$ of a vorton without surrounding electrons, Ref.~\cite{Bedaque:2011fg} makes several approximations.  The first is that instead of computing the $F$ directly for a vortex loop, instead the tension $\mathcal{F}_{\perp}$ for a straight superconducting vortex is used and simply multiplied by the length of the vortex $2\pi R$, which directly neglects curvature effects.  This contribution to $F$ accounts for all of the space in a torus whose radius and thickness are both $R$.  To that contribution, the electromagnetic energy is approximated by its leading multipole contributions--the electric field by the monopole contribution and the magnetic field by the dipole contribution.  These far-field energies account for the space outside a circumscribed sphere of radius $2R$ around the inner torus.  The apple-core-shaped space outside of the torus but inside of the sphere was left unaccounted for.  Artificially partitioning space and only taking only the leading multipole moments both reduce the accuracy of $F$.

The leading far-field electric part of $F$ is easily computed and is given by
\begin{equation}
	F_{E\text{-far}} = \int_{2R}^{\infty} d^{3}r\ \half \left(	\frac{Q}{4\pi r^{2}}	\right)^{2} = \frac{(e Z)^{2}}{16\pi R}
\end{equation}
The electric part of $F$ that is outside of the vortex core but within the torus is given by
\begin{equation}
	F_{E\text{-near}} = 2\pi R \int_{\delta}^{R} d^{2}r_{\perp}\ \half \left(	\frac{\lambda}{2\pi r}	\right)^{2} = \frac{(e Z)^{2}}{8\pi R}\log\left(	\frac{R}{\delta}	\right)
\end{equation}
In order improve upon the results in \cite{Bedaque:2011fg}, we will minimize a modified $F$.  We will minimize
\begin{equation}\label{eq:energyfunction}
	F = F_{\text{old}} - F_{E\text{-far}} - F_{E\text{-near}} + F_{TF}
\end{equation}
where $F_{\text{old}}$ is the main analytical result of \cite{Bedaque:2011fg} and $F_{TF}$ is the energy in the electric fields that we find numerically by solving the Thomas-Fermi equations for a toroidal charge of radius $R$, thickness $\delta$, and charge $eZ$.  In the next section, we will use the Thomas-Fermi approximation to find the semiclassical shape of the electron cloud surrounding a toroidal charge.  Replacing our approximations with $F_{TF}$ accounts for electrons, avoids partitioning space artificially for the electric energy, and avoids the multipole approximation for the electric energy.  We will not change the expressions for the angular momentum or other expressions which depend on the electric field.  For details on the non-Coulomb part of the vorton energy we point the reader to Ref.~(\cite{Bedaque:2011fg}).

\section{Shielding by Electrons} 
\label{sec:shielding_by_electrons}

In this section we give the method for numerically calculating the energy in the electric fields $F_{TF}$, accounting for electrons.  The Thomas-Fermi approach approximates the electron density as
\begin{equation}\label{eq:density}
	n= \oneover{3\pi^{2}}\pF^{3}.
\end{equation}
where $\pF$ is a local fermi momentum and we work in units where $\hbar=c=\epsilon_{0}=1$.  Gauss' law,
\begin{equation}\label{eq:gauss}
	\grad^{2}\phi = e n
\end{equation}
tells us how to compute the electric potential $\phi$ given a fixed distribution of charges $n$.  We also include the equilibrium condition that the marginal energy cost of moving an electron from one point to another should zero,
\begin{equation}
	d\left(	p_{F} - e\phi	\right) = 0
\end{equation} where we have used the ultrarelativistic approximation $\sqrt{p^{2}+m^{2}}\approx \pF$, as we will find that the electrons' fermi momentum is significantly higher than their mass.  This approximation may not be justified for electrons near the edge of the electron cloud, but we will find that most electrons live near the vorton, and thus are contained in a volume $\orderof{10^{6}\text{fm}^{3}}$, meaning that  (for $Z\sim10^{5}$) they have a Fermi momentum \mbox{$\sim 300$ MeV $\gg m_{e} = .511$ MeV} and the ultrarelativistic approximation is appropriate.  We then know that
\begin{equation}\label{eq:cost}
	p_{F} - e\phi = -e\phi_{\infty}
\end{equation}
where $\phi_{\infty}$ is a constant that describes the electric potential at infinity.

Resolving \eqref{eq:density}, \eqref{eq:gauss}, and \eqref{eq:cost} for an equation depending only on $\phi$ yields
\begin{equation}\label{eq:thomas-fermi}
	\grad^{2}\phi = \frac{16}{3}\alpha^{2}(\phi-\phi_{\infty})^{3}
\end{equation}
Since $\phi_{\infty}$ is a constant, $\grad^{2}\phi_{\infty}$ is zero and may be subtracted from the left-hand side, and we can absorb $\phi_{\infty}$ into $\phi$ by a simple change of variables.  Equivalently, we can pick $\phi_{\infty}=0$.  This is a manifestation of gauge invariance, and picking $\phi_{\infty}=0$ is a choice of gauge.  
Eq.~(\ref{eq:thomas-fermi}) is the ultrarelativistic generalization of the usual Thomas-Fermi equation and its solution, with appropriate boundary conditions, is the central theme of this paper.
The Thomas-Fermi approximation neglects the interaction among the electrons but, at the high electron densities occurring in most of the electron cloud this is a very small effect. It also neglects the energy associated to gradients of the density. Again, these gradients are negligible compared to the momenta of the electrons themselves.
By solving this nonlinear differential equation with $\phi_{\infty}=0$ subject to boundary conditions that account for the toroidal charge representing the vorton and the total charge neutrality of the whole system we find a $\phi$ that self-consistently dictates and accounts for the distribution of electrons.

To solve take advantage of the geometry of the situation, we solved \eqref{eq:thomas-fermi} in toroidal coordinates , defined in terms of the usual Cartesian coordinates by
\begin{align}
	u &= -2 \text{ Im }	\arccoth\left(	(\sqrt{x^{2}+y^{2}}+i z)/a	\right)	\nonumber\\
	v &= 2 \text{ Re }	\arccoth\left(	(\sqrt{x^{2}+y^{2}}+i z)/a	\right)	\\
	\varphi &= \arctan(y/x)													\nonumber
\end{align}
where $a$ is the radius of the reference circle (which does not exactly correspond to the radius of the vorton) and Re and Im give the real and imaginary parts, respectively.  We use the fundamental domain
\begin{equation}
	-\pi<u\leq\pi	\hspace{2em} 0\leq v < \infty	\hspace{2em} -\pi < \varphi \leq \pi,
\end{equation}
where a large $v$ corresponds to being near the reference circle, and $v$ near 0 corresponds to a very large circle that comes close to the $z$-axis.  The surface of the vorton occurs at $v_{max}=\sinh\inverse\sqrt{(R/\delta)^{2}-1}$.  Our expression for the energy of a vorton already accounts for the energy in the superconducting core $v>v_{max}$, so we restrict to the domain $0\leq v< v_{max}$.  Excluding electrons from $v>v_{max}$ creates an artificial pressure which can make $\delta$ artificially smaller.  Numerically we find that this effect makes  changes to $\delta$ that are $\sim$10\%. Since our problem has cylindrical symmetry, the ground-state electron distribution will not depend on the axial coordinate $\varphi$ and we suppress it henceforth.

The Laplacian for these $u,v$ toroidal coordinates is
\begin{align}
	\grad^{2}\phi &= \frac{1}{a^2}(\cos(u)-\cosh(v)) \times \nonumber\\
		&\left[(\cos(u) \coth(v)-\text{csch}(v)) \partial_{v}\phi\vphantom{\partial_{v}^{2}} +\sin(u) \partial_{u}\phi\right.\\
		&\left.+(\cos(u)-\cosh(v)) \left(\partial_{u}^{2}+\partial_{v}^{2}\right)\phi\right].\nonumber
\end{align}
We discretized $u$ and $v$ onto a $32\times 32$ grid, and promoted $\phi$ to $\phi(t)$ where $t$ is a fictitious time coordinate.  Then, we numerically solve the coupled ordinary differential equation,
\begin{equation}\label{eq:relaxation}
		\grad^{2}\phi - \frac{16}{3}\alpha^{3}\phi^{3} = \frac{\csch(v)}{(r^{2}-\delta^{2})^{3/2}} \frac{d\phi}{dt}
\end{equation}
 so that if $\phi(t_{f})$ reaches a temporal fixed point at some late time $t_{f}$, it must also automatically satisfy \eqref{eq:thomas-fermi}.  The function multiplying $d\phi/dt$ was picked to alleviate some numerical instabilities, but the final solution of \eqref{eq:thomas-fermi} is independent of this choice as long as $d\phi/dt$ vanishes.

The toroidal coordinate $u$ spans a range $-\pi$ to $\pi$.  The value $u=0$ corresponds to the $xy-$plane with \mbox{$x^{2}+y^{2}>a^{2}$}, while $u=\pi$ corresponds to that same plane but with $x^{2}+y^{2}<a^{2}$.  Since our physical setup has parity across this plane, we can reduce the region where we need to solve \eqref{eq:thomas-fermi} to $0\leq u\leq \pi$, with the boundary conditions
\begin{equation}
	\partial_{u}\phi(0,v) = \partial_{u}\phi(\pi,v) = 0.
\end{equation}
This reduction of the physical space allows us to get a more reliable solution while using a grid of the same size.

Axial symmetry about the $z-$axis (which corresponds to $v=0$) entails that the solution must obey
\begin{equation}
	\partial_{v}\phi(u,0) = 0.
\end{equation}
By symmetry alone we have specified the conditions on three out of the four boundaries of the region.
 
The final boundary is at the wire.  There, we enforce that the electric field is that of a wire of linear charge density $Ze/2\pi R$ and circular cross-section $\pi\delta^{2}$,
\begin{equation}
	\partial_{v}\phi(u,v_{max}) = 2\left(	\frac{Z e}{2\pi R}	\right) \coth\left(	v_{max}	\right).
\end{equation}
This condition matches the electric field to the electromagnetic gauge field inside the vorton, whose form are found explicitly in \cite{Bedaque:2011fg}.
All the boundaries are subject to Neumann conditions (as expected by gauge symmetry), so to numerically stabilize the solution we specify the gauge $\phi(0,0)=\phi_{\infty}=0$.

The important question now arises: how does the system know how many electrons to use to shield the vorton at long distances?  That is, how does it come about that the long-distance field behavior does not include a monopole piece?  It isn't clear that we have implemented this physically desirable constraint.  We will now show that with the symmetries we have specified, toroidal coordinates implement this constraint automatically.

A straightforward manipulation shows that the spherical coordinate $r$ is given by
\begin{equation}
	r^{2} = a^{2} \frac{\cosh v + \cos u}{\cosh v - \cos u}
\end{equation}
and so large $r^{2}$ corresponds to small $u^{2}+v^{2}$.  The multipole expansion of $\phi$,
\begin{equation}
	\phi(\vec{r}) = \phi_{\infty} - \frac{Q_{total}}{4\pi}\frac{1}{r} - \frac{\vec{d}\cdot\hat{r}}{4\pi}\frac{1}{r^{2}} + \cdots
\end{equation}
corresponds to
\begin{equation}
	\phi(u,v) = \phi_{\infty} - \frac{Q_{total}}{4\pi}\frac{\sqrt{u^{2}+v^{2}}}{2a} - \frac{\vec{d}\cdot\hat{r}}{4\pi}\frac{u^{2}+v^{2}}{4a^{2}} + \cdots.
\end{equation}
when both $u$ and $v$ are small.  We choose $\phi_{\infty}$ to be zero as a choice of gauge, and $\vec{d}=0$ is enforced by parity and axial symmetries.  However, it is not obvious that we have enforced $Q_{total}=0$.  Part of the puzzle is that all of the surface at infinity is mapped to a single point, $(u,v)=(0,0)$.  However, on the boundaries $u=0$ and $v=0$ we have already recognized that parity and axial symmetry require $\partial_{u}\phi(0,v)=0$ and $\partial_{v}\phi(u,0)=0$ respectively, meaning that near $(0,0)$ $\phi$ is flat in both directions---exactly what we need to guarantee $Q_{total}=0$.  Charge neutrality is accidentally achieved by our other symmetry conditions in these coordinates, which makes implementing such a solution easy.

\begin{figure}[htbp]
	\centering
		\includegraphics[width=\columnwidth]{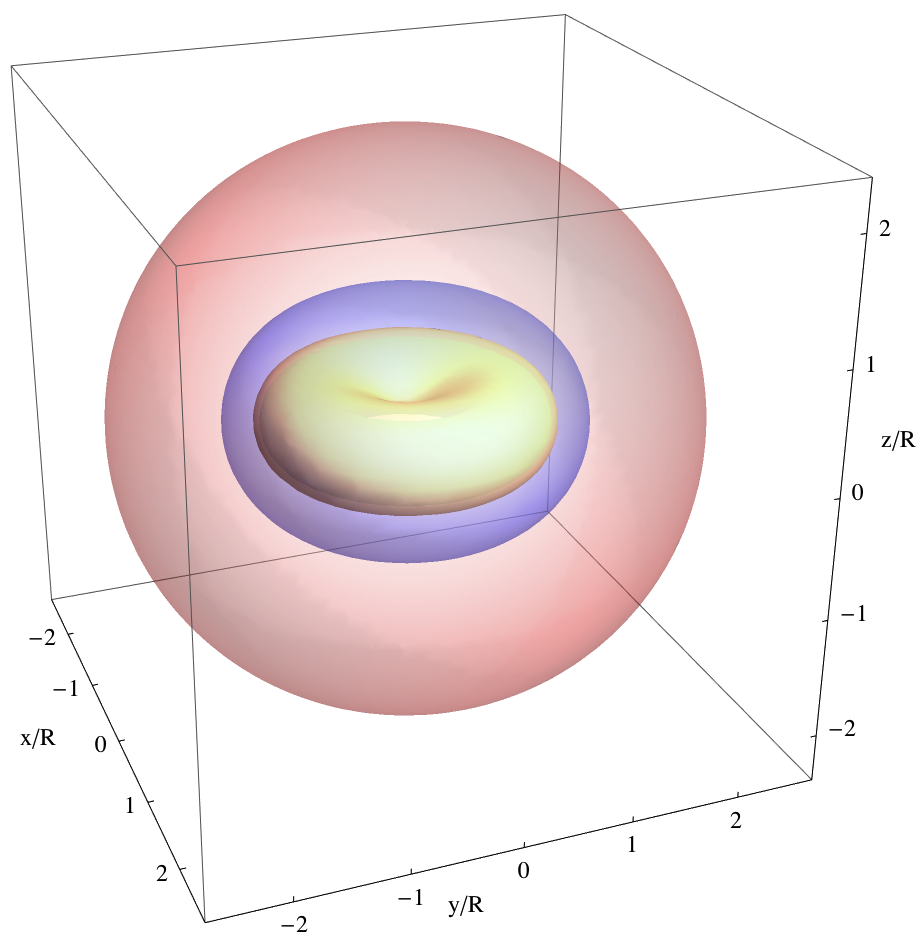}
		\caption{(color online)  Four equipotentials from a typical numerical solution to the Thomas-Fermi equation with a toroidal charge.\label{fig:equipotentials} }
\end{figure}

Unlike the usual application of the  Thomas-Fermi equation to atoms with spherical symmetry which depend only on the charge $Ze$, a dimensionless number, the solution for a toroidal charge does not have a simple scaling behavior, as there are length scale $R$, $\delta$ and not just $Z$ and the electron mass, and the Laplacian does not scale simply as the familiar spherical case does.  Thus, for every $Z$, we solve the differential equation for many $R$ and $\delta$.  It is possible to reduce this three-parameter space to a two-parameter space but for pedagogical simplicity we simply sample the three-dimensional space.  We show one set of equipotentials from a numerical solution in \figref{fig:equipotentials}.  One can easily verify that far from the charge the equipotentials appear spherical, slightly closer the charge the equipotentials resemble oblate spheroids, near the charge there are bialy-shaped equipotentials, and most near the toroidal charge the equipotentials seem toroidal.

 As a check on the numerical accuracy of the procedure, we integrate up the total electric charge outside of the toroidal charge $Z$ and hope to find $-Z$.  That is we ask
\begin{equation}\label{eq:chargeneutrality}
	-Z \stackrel{?}{=} \int_{\text{outside}}2\pi dA\ \grad^{2}\phi(u,v)
\end{equation}
where the $2\pi$ comes from the axial integration and $dA$ is the $uv-$area differential, including the appropriate Jacobian.  This expectation was satisfied to within 1.8\% for every choice of parameters; only 2\% of the numerical solutions had a disagreement between the left- and right-hand sides of more than 1\%; only a quarter of the solutions disagreed by more than 0.5\%.  Moreover, with a finer and finer grid the fulfilling of \eqref{eq:chargeneutrality} is better and better.

From the solution $\phi(u,v)$ for a given $Z$, $R$, and $\delta$, we compute the energy
\begin{equation}
	F_{TF}(Z,R,\delta) = \int 2\pi dA\ \half (-\grad \phi)^{2}.
\end{equation}
For ease of comparison with the results in \cite{Bedaque:2011fg} we computed $F_{TF}$ for $R\in[80,800]$fm in steps of 20 fm, $\delta\in[1,35]$fm with steps of 1 fm, and \mbox{$Z\in\{250,500,750\}\cup([1250,50000]\text{ with steps of 1250})$} and used these data points to construct an interpolating function that worked for any set of parameters within those ranges and extrapolated to larger $\delta$.   Unlike the parts of $F$ which care about the chemical potential and other parameters which might change depending on the environment, $F_{TF}$ only cares indirectly through $R$ and $\delta$.  Therefore, once $F_{TF}(Z,R,\delta)$ is constructed, it can be used for any choice of the chemical potential, the gap, etc.

\begin{figure}[htbp]
	\centering
		\includegraphics[width=\columnwidth]{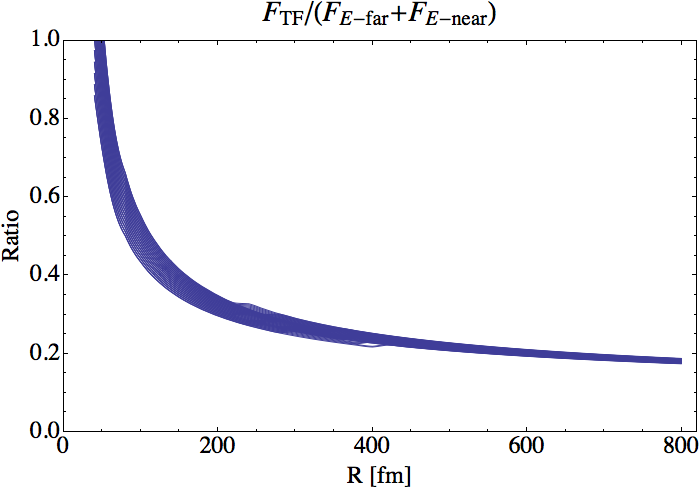}
	\caption{The ratio between the numerically computed energy $F_{TF}$ and the approximation $F_{E\text{-far}}+F_{E\text{-near}}$ as a function of the vorton radius, for $Z\in[1250,40\ 000]$ with steps of 1250 and $\delta\in[20,35]$fm with steps of 1 fm.  The exact identity of each curve is irrelevant; the generic behavior gives rough insight.}
	\label{fig:F-ratio}
\end{figure}

As a way to check the quality of the electron-free approximation, we can examine the ratio of the actual energy in the Thomas-Fermi calculation to the approximated energy, $F_{TF}/(F_{E\text{-far}} + F_{E\text{-near}})$.  We find that for small radii ($R\lesssim 100$fm), where the electrons can't do much shielding, the approximation overestimates by at least a factor of 2.  At very small radii $R\sim\delta$, the electron-free approximation may actually be an underestimate, but the charge ceases to have an obviously toroidal shape, and as other approximations involved in finding the equilibrium radius of a vorton breakdown, is not of physical interest.  With larger radii, where the electrons can shield the Coulomb energy well, the approximation tends to be off by roughly a factor of 5, but the ratio seems to lose any strong $R$ dependence.  This comparison is depicted in \figref{fig:F-ratio}.

\begin{figure}[tbp]
	\centering
		\includegraphics[width=\columnwidth]{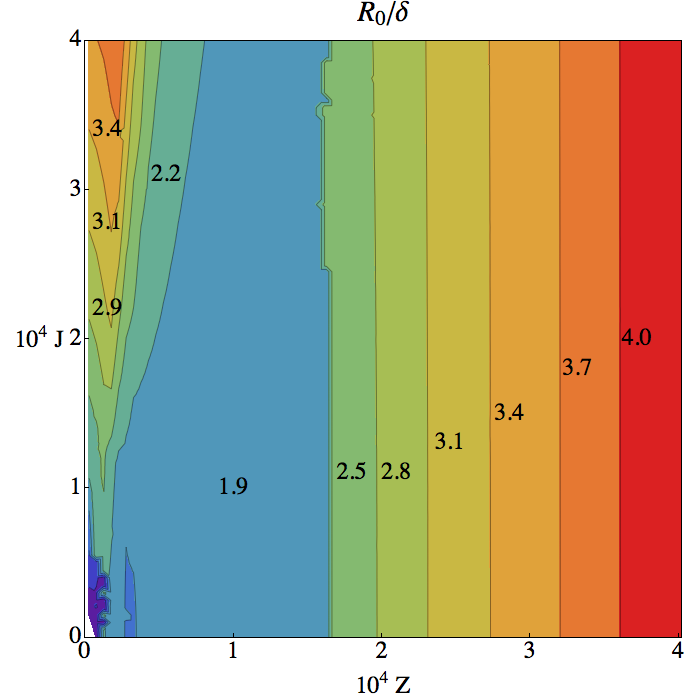}
		\caption{The equilibrium radius $R_{0}$ compared to the string thickness $\delta$ of a shielded vorton at equilibrium as a function of the vorton charge $Ze$ and angular momentum $J$.			\label{fig:Rbydelta}}
\end{figure}
\begin{figure}[tbph]
	\centering
		\includegraphics[width=\columnwidth]{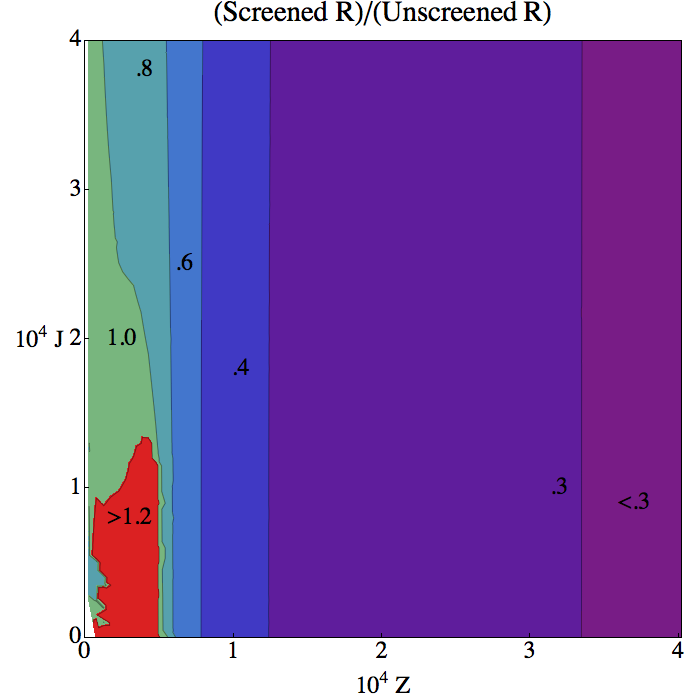}
		\caption{The ratio $R/\delta$ for a vorton with electron shielding compared to $R/\delta$ for the same vorton with no electron shielding as a function of $Z$ and $J$.	The rough nature of the small $Z$ and $J$ region is due to sampling a rough grid---a finer grid would resolve these features.		\label{fig:compare}}
\end{figure}

With $F_{TF}(Z,R,\delta)$ in hand, we can pick a $(Z,J)$ pair, and simply minimize the energy $F$ given in \eqref{eq:energyfunction}.  For concreteness and ease of comparing to unshielded vortons, we use the same example parameters as \cite{Bedaque:2011fg}, a gap $\Delta=66$ MeV, an overall chemical potential $\mu=450$ MeV, and quark masses $(m_{u}, m_{d}, m_{s})\approx(2,5,95)$ MeV.  These parameters don't appear in $F_{TF}$ and thus one calculation of $F_{TF}$ empowers us to examine any relevant density and gap.  We show $R/\delta$ as a function of $(Z,J)$ in \figref{fig:Rbydelta}, and we compare the equilibrium radii of shielded and unshielded vortons in \figref{fig:compare}.  With quick examination of \figref{fig:Rbydelta} it is evident that the effect of electron shielding shrinks the equilibrium radius for values of $(Z,J)$ where the unshielded calculation is believable ($Z\gtrsim 6000$), and that this effect becomes stronger for larger $Z$.  

One interesting effect that can be seen in \figref{fig:compare} is that shielded vortons are actually larger than their unshielded counterparts for small $Z$ and $J$.  This behavior fits with the low-$R$ trend seen in \figref{fig:F-ratio} and should be interpreted as the electron degeneracy helping prevent the vorton's collapse.  However, the results in this regime are not trustable, as it is precisely the same region where the fictitious pressure that arises from restricting the electrons to $v>v_{max}$ shrinks $\delta$ by 20\% to 40\%.  To understand this area well, one would have to allow electrons into the excluded regime in $v$, and would need to solve the Thomas-Fermi equation \eqref{eq:thomas-fermi} simultaneously with the equations of motion for the two kaon types, which is beyond the scope of this work.

More careful examination of \figref{fig:Rbydelta} shows that not only are shielded vortons smaller, but compared to unshielded vortons they grow more slowly as one increases $Z$.  This is the expected behavior, as to first order putting a positive charge in the vorton is cancelled by a shielding electron.  If this cancellation were exact, a shielded vorton's size would be independent of $Z$. Only when one includes Pauli blocking does a shielded vorton's equilibrium radius grow.  A related final observation is the increasing success of the electric shielding by electrons.  As $Z$ gets larger, the ratio of shielded equilibrium radius and the unshielded equilibrium radius drops.  This is also expected, as larger $Z$ corresponds to larger $R$, and a larger $R$ enables electrons to live in the central hole of the vorton more readily, shielding the vorton's charge more effectively.

\section{Summary and Outlook} 
\label{sec:summary_and_outlook}

In this paper we consider the effect of electron shielding on the equilibrium radius of vortons in the \CFLKz\ phase at high density.  For parameter values where unshielded vortons were reasonably large and had a clear toroidal shape with an aspect ratio $R/\delta$ of a few ($Z\gtrsim 6000$), the electron shielding of vortonium, as expected, shrinks both the radius and aspect ratio for a given $Z$ and also diminishes the response of $R/\delta$ to a change in $Z$.

By allowing that portion to be shielded with electrons and applying a relativistic Thomas-Fermi approximation, we have accounted for the electron cloud surrounding such a vorton and have relieved some of uncontrolled approximations concerning the toroidal geometry of vortons in \CFLKz.  We have shown that the energy in the electric field considered in Ref.~\cite{Bedaque:2011fg} is unrealistically optimistic insofar as it drastically changes the equilibrium shape of the vorton as compared to the more realistic vortonium.  The reduced size of vortonium may be understood by realizing that the far-fields considered for unshielded vortons is predominantly a monopole piece while for vortonium it is a quadrupole.

Two directions seem obvious for pursuing the physics of \CFLKz\ vortons further.  The first is to try to cure all of the issues we have touched on in this work:  attempt to find, in a toroidal geometry, the actual mean-field wavefunctions of the neutral and charged kaon condensates for a vorton, while simultaneously incorporating electron shielding and allowing electrons to penetrate the core and not artificially excluding them, and could account for curvature effects.  Work in this direction will almost certainly require intensive numerical calculations.

The other possible direction is to try to estimate the vorton production rate  in the early times in the neutron star life as it goes from a high temperature disordered phase to a low temperature \CFLKz\ phase. This number is important to gauge the vorton influence on neutron star phenomenology.   Along the same lines, one might consider the decay rate of these objects through weak effects, to try and estimate how long any created population might persist in these extreme environments.  This line of inquiry would help prioritize any other work on vortons at high density, and would provide the most direct route to understanding if vortons might contribute to observable astrophysical phenomena.
\acknowledgments

P.~F.~B. and E.~B.\ are supported by the U.S. Department of Energy under grant \#DE-FG02-93ER-40762.  E.~B. also thanks Jefferson Science Associates for support through the JSA/JLab Graduate Fellowship program.   N.N. was encouraged by the summer research program at Montgomery Blair High School.

\bibliography{vortonium}
\end{document}